\documentclass[12pt]{article}

\usepackage{epsf}
\usepackage{epsfig}
\usepackage{cite}
\usepackage{amsmath}
\usepackage{graphics}

\begin{document}

\setlength{\textheight}{21.5cm}
\setlength{\oddsidemargin}{0.cm}
\setlength{\evensidemargin}{0.cm}
\setlength{\topmargin}{0.cm}
\setlength{\footskip}{1cm}
\setlength{\arraycolsep}{2pt}

\renewcommand{\thefootnote}{\#\arabic{footnote}}
\setcounter{footnote}{0}

\newcommand{\gtrsim}{ \mathop{}_{\textstyle \sim}^{\textstyle >} }
\newcommand{\lesssim}{ \mathop{}_{\textstyle \sim}^{\textstyle <} }
\newcommand{\rem}[1]{{\bf #1}}
\renewcommand{\thefootnote}{\fnsymbol{footnote}}
\setcounter{footnote}{0}
\def\thefootnote{\fnsymbol{footnote}}

\hfill October 2009\\

\hfill IPMU09-0121\\

\vskip .5in

\begin{center}

\bigskip
\bigskip

{\Large \bf Number and Entropy of Halo Black Holes} 

\vskip .45in

{\bf Paul H. Frampton\footnote{frampton@physics.unc.edu}
$^{(a,b)}$
and Kevin Ludwick\footnote{kludwick@physics.unc.edu}$^{(a)}$}

\vskip .3in

{\it $^{(a)}$ Department of Physics and Astronomy, University of North Carolina,
Chapel Hill, NC 27599-3255.}

{\it $^{(b)}$ Institute for the Physics and Mathematics of the
Universe, University of Tokyo, Kashiwa, Chiba 277-8568, Japan.}

\end{center}

\vskip .4in 
\begin{abstract}
Based on constraints from microlensing and disk stability, both with and without
limitations from wide binary surveys, we estimate the total
number and entropy of intermediate mass black holes.
Given the visible universe comprises $10^{11}$
halos each of mass $\sim 10^{12} M_{\odot}$, typical core
black holes of mean mass $\sim 10^7 M_{\odot}$
set the dimensionless  entropy ($S/k$)
of the universe at a thousand googols.
Identification of all dark matter as 
black holes sets the dimensionless entropy of the universe at 
ten million googols, implying that dark matter
can contribute over $99\%$ of entropy, which favors all
dark matter as black holes in the mass regime of $ \sim 10^{5} M_{\odot}$.
\end{abstract}

\renewcommand{\thepage}{\arabic{page}}
\setcounter{page}{1}
\renewcommand{\thefootnote}{\#\arabic{footnote}}

\newpage

\noindent {\it Introduction}

\bigskip

\noindent The identification of dark matter, for which there is compelling
evidence from its gravitational effects in galaxies and clusters thereof,
is an important outstanding question. Dark matter makes up some eighty percent
of matter and a quarter of the energy content of the universe. If 
dark matter is composed of only one constituent, its mass
is uncertain by almost eighty orders of magnitude from
$10^{-15}$ to $10^{+65}$ GeV.

\bigskip

\noindent Of course, it would be reassuring to identify dark matter
by its production in particle colliders and by its detection in
terrestrial experiments. On the other hand, the dark matter
constituent may equally be, as assumed here, in a completely
different and collider-inaccessible mass regime heavier than the Sun.

\bigskip

\noindent The observational limits on the occurrence of such multi-solar mass
astrophysical objects in the halo have considerably changed recently.
There remain microlensing limits \cite{ml1,ml2} on masses
below a solar mass and slightly above. There are also respected limits
from numerical study \cite{ds1} of disk stability at
ten million solar masses and slightly below.

\bigskip

\noindent For the intermediate mass region, a possible constriant
comes from the occurrence of gravitationally bound
binary stars at high separation approaching one parsec.
Here the situation has changed recently, and the
bounds are far more relaxed, possibly non-existent.
The first such analysis \cite{Yoo} allowed only some
ten percent of halo dark matter for most
of the mass range. A more recent analysis \cite{Quinn}
permits fifty perecent and cautions that the sample
of binaries may be too small to draw any solid conclusions.

\bigskip

\noindent In the following, we adopt constraints from
microlensing and disk stability but keep an
open mind with respect to the wide binaries. We estimate
the total number and total entropy
of the black holes per halo and hence (simply multiplying by $10^{11}$)
in the universe, assuming as in \cite{PHF2009} that all dark
matter can be identified as black holes.

\bigskip

\noindent {\it Number of Black Holes}

\bigskip

\noindent To estimate number and subsequently entropy
of black holes we simplify by taking as possible masses
$10^n M_{\odot}$ with $n$ integer, $1 \leq n \leq 7$.
Further, we assume the constraints from wide binaries
\cite{Quinn} for different $n$ are independent of each other.

\bigskip

\noindent We make our analysis first
with binary constraints, denoted simply as ``with",
then with no binary constraints, denoted as ``without". 
Let $f_n$ be the fraction of the halo dark matter
composed of mass $10^n M_{\odot}$ black holes. The total halo mass is
taken to be $10^{12} M_{\odot}$ whereupon

\begin{equation}
\Sigma_n f_n = 1
\label{fn}
\end{equation}
and the number $N_n$ is
\begin{equation}
N_n = f_n 10^{12-n}.
\label{Nn}
\end{equation}

\bigskip

\noindent The ``with" constraints on the $f_n$ are

\begin{eqnarray}
0 & \leq & f_1 \leq 0.4 \nonumber \\
0 & \leq & f_2 \leq 1.0 \nonumber \\
0 & \leq & f_3 \leq 0.5 \nonumber \\
0 & \leq & f_{4,5,6} \leq 0.4 \nonumber \\
0 & \leq & f_7 \leq 0.3.
\label{with}
\end{eqnarray}

\bigskip

\noindent For the ``without" constraints, the $f_{1,2,7}$ ranges
remain unchanged while the $f_{3,4,5,6}$ are free, namely

\begin{eqnarray}
0 & \leq & f_1 \leq 0.4 \nonumber \\
0 & \leq & f_{2,3,4,5,6} \leq 1.0 \nonumber \\
0 & \leq & f_7 \leq 0.3.
\end{eqnarray}

\bigskip

\noindent Allowing the $f_n$ to vary by increments $\Delta f_n = 0.1$
for $1 \leq f_n \leq (f_n)_{max}$ we allow the black holes
to have $\nu$ different mass (or $n$) values with $1 \leq \nu \leq 7$.
For the ``with" constraints, we then find numbers of black holes as follows:

\bigskip
\bigskip

\begin{center}
\bigskip

{\bf Table 1}

\bigskip

\begin{tabular}{||c||c|c|c|c|c||}
\hline\hline
$\nu$  & \# choices   & $N_{mean}$ & $N_{median}$ & $N_{max}$ & $N_{min}$ \\ 
\hline\hline
1 & 1 & $1.0 \times 10^{10}$ & $1.0 \times 10^{10}$ & 
$1.0 \times 10^{10}$ & $1.0 \times 10^{10}$    \\
\hline
2 & 24 & $1.2 \times 10^{10}$ & $8.0 \times 10^9$ & 
$4.6 \times 10^{10}$ & $5.5 \times 10^9$  \\
\hline
3 & 365 & $1.4 \times 10^{10}$ & $6.1 \times 10^9$ & 
$4.5 \times 10^{10}$ & $3.4 \times 10^6$   \\
\hline
4 & 1660 & $1.6 \times 10^{10}$ & $1.2 \times 10^{10}$ & 
$4.4 \times 10^{10}$ & $1.2 \times 10^7$   \\
\hline
5 & 2106 & $1.6 \times 10^{10}$ & $1.3 \times 10^{10}$ & 
$4.3 \times 10^{10}$ & $1.1 \times 10^8$   \\
\hline
6 & 822 & $1.6 \times 10^{10}$ & $1.2 \times 10^{10}$ & 
$4.2 \times 10^{10}$ & $1.1 \times 10^9$   \\
\hline
7 & 83 & $1.6 \times 10^{10}$ & $1.2 \times 10^{10}$ & 
$4.1 \times 10^{10}$ & $1.1 \times 10^{10}$   \\
\hline\hline
\end{tabular}

\end{center}

\bigskip

\noindent For the ``without" case we find:

\bigskip
\bigskip

\begin{center}

\bigskip

{\bf Table 2}

\bigskip

\begin{tabular}{||c||c|c|c|c|c||}
\hline\hline
$\nu$  & \# choices  & $N_{mean}$ & $N_{median}$ & $N_{max}$ & $N_{min}$ \\ 
\hline\hline
1 & 5 & $2.2 \times 10^9$ & $1.0 \times 10^9$ & 
$1.0 \times 10^{10}$ & $1.0 \times 10^6$  \\
\hline
2 & 125 & $6.1 \times 10^9$ & $9.0 \times 10^8$ & 
$ 4.6 \times 10^{10}$ & $7.3 \times 10^5$  \\
\hline
3 & 890 & $1.0 \times 10^{10}$ & $4.0 \times 10^9$ & 
$ 4.5 \times 10^{10}$ & $1.6 \times 10^6$   \\
\hline
4 & 2340 & $1.4 \times 10^{10}$ & $1.0 \times 10^{10}$ & 
$4.4 \times 10^{10}$ & $1.1 \times 10^7$   \\
\hline
5 & 2346 & $1.5 \times 10^{10}$ & $1.2 \times 10^{10}$ & 
$4.3 \times 10^{10}$ & $1.1 \times 10^8$    \\
\hline
6 & 840 & $1.6 \times 10^{10}$ & $1.2 \times 10^{10}$ &
$4.2 \times 10^{10}$ & $1.1 \times 10^9$    \\
\hline
7 & 83 & $1.6 \times 10^{10}$ & $1.2 \times 10^{10}$ & 
$4.1 \times 10^{10}$ & $1.1 \times 10^{10}$   \\
\hline\hline
\end{tabular}

\end{center}

\bigskip
\bigskip

\noindent Study of Tables 1 and 2 reveals a number of things about
the putative intermediate mass black holes which may dominate
the matter content. First, the comparison of the tables reveals that
the wide binary constraints, as they stand, do not affect
the numbers very much. Thus, unless and until a much bigger
sample of wide binaries is found (if they exist), the conclusions
about numbers of black holes
in a halo is insentive to their consideration.

\bigskip

\noindent As expected from the defining formula, Eq. (\ref{Nn}),
the number of black holes per halo can range from about a million
to a few times ten billion. By sampling 
distributions of the masses, not just a single mass, 
Tables 1 and 2 reveal that
the most likely number is at the high end, close to
ten billion per halo.

\bigskip

\noindent Since there are generically $10^{11}$ halos,
this implies a total number of about a
billion trillion black holes in the universe.

\bigskip
\bigskip
\bigskip

\noindent {\it Entropy of Black Holes}

\bigskip
\bigskip

\noindent We can similarly estimate the total entropy
of the halo black holes by exploiting the BPH entropy
formula \cite{P,B,H}, which says that for a black hole
with mass $M_{BH} = \eta M_{\odot}$, the entropy is $S_{BH} = 10^{78} \eta^2$.

\bigskip

\noindent For the ``with" case, this gives the numbers
for halo entropy

\bigskip
\bigskip

\begin{center}

\bigskip

{\bf Table 3}

\bigskip

\begin{tabular}{||c||c|c|c|c|c||}
\hline\hline
$\nu$  & \# choices  & $S_{mean}$ & $S_{median}$ & $S_{max}$ & $S_{min}$ \\ 
\hline\hline
1 & 1 & $1.0 \times 10^{92}$ & $1.0 \times 10^{92}$ & 
$1.0 \times 10^{92}$ &  $1.0 \times 10^{92}$ \\
\hline
2 & 24 & $3.0 \times 10^{95}$& $7.1 \times 10^{93}$ & 
$3.0 \times 10^{96}$ & $6.4 \times 10^{91}$   \\
\hline
3 & 365 & $7.5 \times 10^{95}$ & $2.2 \times 19^{95}$ & 
$3.4 \times 10^{96}$ & $1.5 \times 10^{92}$   \\
\hline
4 & 1660 & $1.1 \times 10^{96}$ & $1.0 \times 10^{96}$ & 
$3.4 \times 10^{96}$ & $1.1 \times 10^{93}$ \\ 
\hline
5 & 2106 & $1.4 \times 10^{96}$ & $ 1.2 \times 10^{96}$ & 
$3.4 \times 10^{96}$ & $1.1 \times 10^{94}$  \\
\hline
6 & 822 & $1.5 \times 10^{96}$ & $1.2 \times 10^{96}$ & 
$3.3 \times 10^{96}$ & $1.1 \times 10^{95}$   \\
\hline
7 & 83 & $1.6 \times 10^{96}$ & $1.2 \times 10^{96}$ & 
$3.2 \times 10^{96}$ & $1.1 \times 10^{96}$  \\
\hline\hline
\end{tabular}

\end{center}

\bigskip
\bigskip

\noindent while for the ``without" case we find

\bigskip
\bigskip

\begin{center}

\bigskip

{\bf Table 4}

\bigskip

\begin{tabular}{||c||c|c|c|c|c||}
\hline\hline
$\nu$  & \# choices  & $S_{mean}$ & $S_{median}$ & $S_{max}$ & $S_{min}$ \\ 
\hline\hline
1 & 5 & $2.2 \times 10^{95}$ & $1.0 \times 10^{95}$ & 
$1.0 \times 10^{96}$  & $1.0 \times 10^{92}$   \\
\hline
2 & 125 & $4.5 \times 10^{95}$ & $8.0 \times 10^{94}$ & 
$3.7 \times 10^{96}$ & $6.4 \times 10^{91}$  \\
\hline
3 & 890 & $7.7 \times 10^{95}$ & $3.0 \times 10^{95}$ & 
$3.6 \times 10^{96}$ & $1.5 \times 10^{92}$  \\ 
\hline
4 & 2340 & $1.1 \times 10^{96}$ & $1.0 \times 10^{96}$ & 
$3.5 \times 10^{96}$ & $1.1 \times 10^{93}$   \\
\hline
5 & 2346 & $1.3 \times 10^{96}$ & $1.2 \times 10^{96}$ & 
$3.4 \times 10^{96}$ & $1.1 \times 10^{94}$  \\
\hline
6 & 840 & $1.5 \times 10^{96}$ & $1.2 \times 10^{96}$ & 
$3.3 \times 10^{96}$ & $1.1 \times 10^{95}$   \\
\hline
7 & 83 & $1.6 \times 10^{96}$ & $1.2 \times 10^{96}$ & 
$3.2 \times 10^{96}$ & $1.1 \times 10^{96}$    \\
\hline\hline
\end{tabular}

\end{center}

\bigskip
\bigskip

\noindent Tables 3 and 4 contain much information germane to the
central idea that dark matter be identified as black holes.

\bigskip

\noindent The biggest known contributor
of black holes in a halo is the core supermassive black
hole (SMBH). In the Milky Way it is Sag A* and for a typical galaxy
a core SMBH has mass $M_{SMBH} \sim 10^7 M_{\odot}$.  Its PBH
entropy is therefore about $\sim 10^{92}$.

\bigskip

\noindent Multiplying by $10^{11}$, the number of halos,
shows that these SMBHs contribute about $10^{103}$, or a thousand 
googols, to
the entropy of the universe as emphasized in \cite{PHF2009}.

\bigskip

\noindent The conventional wisdom is that the SMBHs are the single
dominant contributor to the entropy
of the universe, which is therefore
about a thousand googols.

\bigskip
\bigskip

\noindent From our Tables 3 and 4 we can arrive at a very different conclusion.

\newpage

\noindent {\it Reconsideration of the Entropy of the Universe}

\bigskip

\noindent Let us take the viewpoint that the universe, by
which we mean the visible universe, is an isolated system
in the usual sense of thermodynamics and statistical mechanics.
In accord with the usual statistical law
of thermodynamics, the entropy of the universe
will increase to its maximum attainable value.

\bigskip

\noindent The natural unit for the dimensionless entropy
of the universe $S/k = \ln \Omega$ is the googol ($10^{100}$).
The supermassive black holes (SMBHs) at galactic cores contribute
about a thousand googols.

\bigskip

\noindent The holographic bound \cite{Hooft}
on information or entropy
contained in a three-volume is that it be not above
the surface area as measured in Planck units $(10^{-33} cm^2)$.
If we take the visible universe to be a sphere of
radius $3 \times 10^{10} ly \sim 3 \times 10^{18} cm$, the
maximum  entropy is $\sim 10^{124}$ or a trillion trillion googols.
This would be the entropy if the universe
were one black hole of mass $10^{23} M_{\odot}$.

\bigskip

\noindent The numbers in Tables 3 and 4 suggest that the entropy
contribution from dark matter can
exceed that of the SMBHs by orders
of magnitude. Taking the view that
increasing total entropy plays a dominant role in cosmological evolution
strongly favors the formation of black holes
in the $10^5 M_{\odot}$ mass range and the view that they
constitute all dark matter.

\bigskip

\noindent We are more confident about the present status of dark matter
than of it detailed history but, of course, an interesting and legitimate
question is: how did the black holes originate? One possible formation
is as remnants of Population-III (henceforth Pop-III) stars
formed at a redshift $Z \sim 25$.
These Pop-III stars are necessary to explain the
metallicity of Pop-I and Pop-II stars that formed later. Such
Pop-III stars are not well understood but we expect they can be very
massive, $10^5 M_{\odot}$, to live for a
short time, less that a million years, then explode leaving
black holes which have a total mass that is a significant
fraction of the original star's mass.
Nevertheless, it is very unlikely \cite{Madau}
that a sufficent number of
Pop-III stars can form to make all dark matter.
Thus, the IMBHs may have formed in the early universe as
primordial black holes.\footnote[1]{Note that the constraints
in \cite{Ostriker} apply at the recombination era
and subsequent black hole mergers can occur.}

\bigskip

\newpage

\noindent The first item of business is therefore
to confirm that there are millions of large black holes
in our halo and in others.

\bigskip

\noindent The ESA Gaia project is planned to survey billions
of stars in our galaxy, the Milky Way, and should
enable obtaining a large sample of gravitationally bound
wide binaries which can be analyzed for evidence
of black holes perturbing them.

\bigskip

\noindent The goal of the SuperMACHO project is to
identify the objects which produced existing microlensing
events and should allow the observation of higher
longevity microlensing signals corresponding to the mass ranges
suggested for the dark matter black holes.

\bigskip

\noindent Finally, if we truncate to $n \leq 5$, since Pop-III
stars or IMBHs with higher masses seem unlikely\cite{Madau}, then the
typical number of IMBHs per halo is $\sim 10^{10}$,
giving about one million googols for the entropy
of the universe. The majority of entropy may be concentrated in
a tiny fraction of the total number of black
holes as can be seen by studying examples,
{\it e.g.}, $f_2=f_5=0.5$.

\bigskip

\noindent The key motivation for our believing this
interpretation of dark matter, as opposed to
an interpretation involving microscopic particles,
comes from consideration of the entropy
of the universe. The SMBHs at galactic cores contribute
about a thousand googols to the overall dimensionless entropy.
As seen in the present article, dark matter in the form of black
holes can contribute as much as a million googols and thus
make up over $99\%$ of the cosmic entropy, which
is sufficient reason, if we adopt that the universe
is an isolated system
to which the second law of thermodynamics
is applicable, for taking it seriously.

\bigskip

\noindent Hopefully future observations will be able
to identify dark matter as black holes.

\newpage

\begin{center}

\section*{Acknowledgements}

\end{center}

This work was supported in part 
by World Premier International Research Center
Initiative (WPI Initiative), MEXT, Japan and
by the U.S. Department of Energy under Grant
No. DE-FG02-06ER41418.

\end{document}